\documentclass[pdflatex,sn-mathphys-num]{sn-jnl}

\usepackage{graphicx}%
\usepackage{multirow}%
\usepackage{amsmath,amssymb,amsfonts}%
\usepackage{amsthm}%
\usepackage{mathrsfs}%
\usepackage[title]{appendix}%
\usepackage{xcolor}%
\usepackage{textcomp}%
\usepackage{manyfoot}%
\usepackage{booktabs}%
\usepackage{algorithm}%
\usepackage{algorithmicx}%
\usepackage{algpseudocode}%
\usepackage{listings}%
\usepackage{siunitx}%
\usepackage{parskip}%
\usepackage{pdflscape}

\theoremstyle{thmstyleone}%
\theoremstyle{thmstyletwo}%

\theoremstyle{thmstylethree}%

\begin{document}

\title[Article Title]{Computational modelling reveals advantages of ventricular over atrial cardiomyocytes to power self-guided biorobots}


\author[1]{\fnm{Nathan} \sur{Lilly}}
\author[2]{\fnm{Guy} \sur{Bewick}}
\equalcont{These authors contributed equally to this work.}

\author[1]{\fnm{Claudiu Vasile} \sur{Giuraniuc}}
\equalcont{These authors contributed equally to this work.}

\author*[3]{\fnm{Maria Elena} \sur{Giannaccini}}\email{mariaelena.giannaccini@nottingham.ac.uk}
\equalcont{These authors contributed equally to this work.}

\affil[1]{\orgdiv{Department of Physics}, \orgname{University of Aberdeen}, \orgaddress{\city{Aberdeen}, \country{United Kingdom}}}

\affil[2]{\orgdiv{Institute of Medical Sciences, School of Medicine, Medical Sciences \& Nutrition}, \orgname{University of Aberdeen}, \orgaddress{\city{Aberdeen}, \country{United Kingdom}}}

\affil[3]{\orgdiv{School of Computer Science}, \orgname{University of Nottingham}, \orgaddress{\city{Nottingham}, \country{United Kingdom}}}



\clearpage
\abstract{Solid tumours and other hypoxic regions are difficult to treat with conventional systemic drug delivery, which distributes therapeutics non-specifically and increases toxicity in off-target sites. Biohybrid microrobots driven by living cells offer a self-powered, bio-compatible solution to autonomous targeted delivery, however, existing platforms often require external stimuli to move and locate towards target zones. We present a computational model of an H-shaped biohybrid microrobot, actuated by neonatal rat cardiomyocytes coupled to the body through an anisotropic friction-ratchet generating a net forward locomotion. The model is simulated under normoxic straight-line conditions and hypoxic gradient-steering conditions for two distinct phenotypes, atrial and ventricular. Ventricular cells produce a $4.35\times$ increase in locomotion speed over atrial cells at an equivalent cost of transport and a $2.91\times$ greater heading response under an applied gradient, confirming that phenotype is an underexploited design variable. The computational model in this study shows autonomous, gradient-based steering in a cardiomyocyte-based platform and the effectiveness of treating cardiomyocyte phenotype as an explicit design parameter. These findings support the long-term feasibility of cardiomyocyte-driven microrobots for autonomous, self-targeted therapeutic delivery in hypoxic regions and as such offer a route to reducing systemic drug exposure relative to conventional treatment. }




\maketitle

\section{Introduction}
Solid tumours and other hypoxic tissues are difficult to treat with conventional systemic drug delivery due to the non-specific distribution of therapeutics throughout the host reducing efficacy at the target site and increasing off-target toxicity \cite{hockel2001}. A robot capable of autonomously navigating hypoxic regions and delivering therapeutics locally would address this limitation directly. Conventional microrobots are poorly suited to such a task as their actuators (pneumatic, magnetic) rely on non-biocompatible materials and typically require external stimuli to move and navigate \cite{lin2022}.

Biohybrid robots, in which living cells provide the actuation, offer a self-powered and inherently bio-compatible alternative that operate at scales directly relevant to biological environments \cite{calisti2017}. Cardiomyocytes emerge as a particularly compelling actuator choice for this purpose as they contract autonomously, negating the need for external stimuli to generate a motive force. Additionally, cardiomyocytes respond to localised oxygen concentration, a property that can be exploited for chemotactic navigation. Cardiomyocyte devices in the literature have demonstrated thin-film locomotion \cite{alford2010} and free-crawling capabilities \cite{sun2020}, and oxygen/light-responsive actuation has been used to steer tissue-engineered swimmers \cite{park2016}. Notably however, existing cardiomyocyte-driven crawlers rely on externally imposed directionality such as magnetic field gradients rather than cardiomyocytes' own environmental sensitivity \cite{sun2020}. Separately, oxygen-gradient (aerotactic) steering has been demonstrated in microorganism-driven biohybrid robots \cite{lin2022} but has yet to be demonstrated for a cardiomyocyte-based platform. Within the literature, no existing platform has been identified that combines autonomous actuation of cardiomyocytes with a self-generated chemotactic steering response to an oxygen gradient and none that have treated cardiomyocyte phenotype as an explicit design variable for locomotion performance. 

This paper addresses that gap. We present a computational model of an H-shaped biohybrid microrobot, constructed from two parallel silicone arms interconnected by a rigid gold rod crossbar to provide electrical continuity for the cardiomyocytes in the arms and actuated by neonatal rat cardiomyocytes coupled to the body through a peristaltic, anisotropic friction-ratchet. This geometry is simulated under both normoxic straight-line and hypoxic gradient-steering conditions, for two cardiomyocyte phenotypes, atrial and ventricular providing a quantitative basis for evaluating phenotype selection and oxygen-gradient steering as design parameters for biohybrid locomotion.

\section{Methods}
\label{sec:system_model}
\subsection{Robot Geometry and Mechanics}
\label{subsec:geometry}
\begin{figure}[t]
  \centering
  \includegraphics[width=0.75\linewidth]{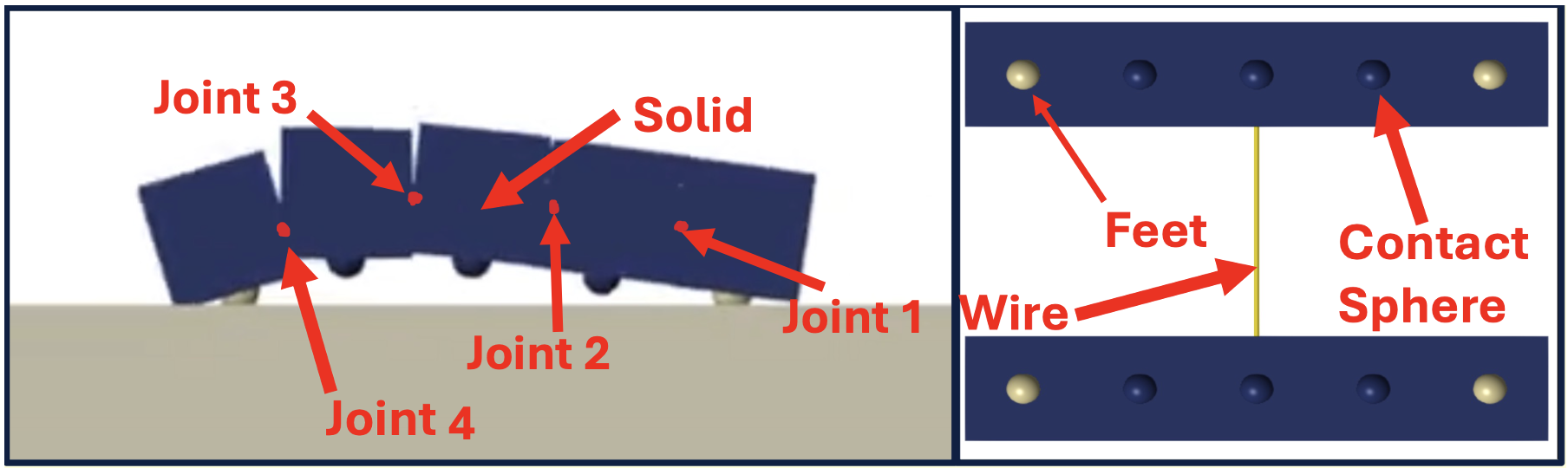}
  \caption{Left: Lateral view of the model during actuation, illustrating the 5 segment discretisation of a single arm and the contact spheres (gold). Right: Ventral view of the full H-shaped robot at rest, showing both arms (blue), the gold crossbar (yellow), the corner contact legs (gold spheres) and mid-arm support spheres (blue spheres)}
    \label{fig:robot_schematic}
\end{figure}

The modelled robot is an H-shaped geometry consisting of two Ecoflex 00-30 beams each with dimensions \qtyproduct{5 x 1 x 1}{\milli\meter} (dimensions chosen as a manufacturing condition) interconnected at their midpoint by a gold rod crossbar of length \SI{3}{\milli\meter} and radius \SI{25}{\micro\meter}. The resulting H-shaped body plan is illustrated schematically in Fig \ref{fig:robot_schematic} and was chosen to break bilateral symmetry allowing for the independent actuation of each arm. This independent actuation of each arm is a mechanical condition required for a differential-torque steering mechanism that a single-arm geometry could not support. Five spherical contact legs ($r = 150$~\textmu m, Ecoflex~00-30) are positioned on the ventral surface of each arm, two corner legs at 1~mm (posterior) and 4~mm (anterior) and three intermediate support hemispheres. The anterior and posterior (corner) legs act as the primary friction-ratchet anchor points whilst the intermediate support hemispheres act as a simulation condition to prevent the midsection from penetrating the simulated surface. Each leg is hemispherical in geometry which aligns with the recommended approach from Mathworks for a multi-contact Simscape Multibody simulation \cite{mathworks2024}. Hemispherical geometry ensures a smooth contact interface between the leg and the surface and allows the contact forces to be computed via the Simscape Spatial Contract Force block, which implements a penalty (spring–damper) model, $F_n = s(d,w)\,(k_c d + b\,\dot{d})$, with contact stiffness $k_c = 2$~N\,m$^{-1}$ and damping $b = 0.5$~N\,s\,m$^{-1}$.

The model uses a lumped-parameter approach discretising each arm into a chain of five equal rigid segments (Dimensions of \qtyproduct{1 x 1 x 1}{\milli\meter}) connected by four revolute joints (J1, J2, J3, J4 Rear-Front), a resolution which balances physical fidelity with solver stability  \cite{contreras2022, mallikarachchi2024}. This approach is an approximation, representing the continuum of the Ecoflex arms as a series of discrete rigid segments linked by revolute joints with joint stiffness representing material compliance. The joint torsional stiffness ($k$) is derived from beam bending theory:

\begin{equation}
     k = \frac{E_{\mathrm{Ecoflex}}I} {L_{\mathrm{seg}}}
     \label{eq:joint_stiffness} 
\end{equation}

Where $E$ is Young's modulus, I is the second moment of area computed as $I = \frac{w\,t^{3}}{12}$ and $L_{seg}$ is the length of one discrete segment.

Joint damping ($c$) is set to $c = 0.05\,k = 2.29 \times 10^{-10}$~N\,m\,s\,rad$^{-1}$ as a numerical stabilisation parameter appropriate for penalty-based micro-scale contact systems \cite{berthold2023, mathworks2024}.
The gold rod is treated as a rigid body throughout, justified by the ${\sim}1.4 \times 10^{6}$-fold difference in Young's modulus ($E$) between gold ($E_{\mathrm{Au}} = 79$~GPa) and Ecoflex ($E_{\mathrm{Ecoflex}} = 55$~kPa).
  
\begin{table}[t]
\caption{Geometric and material parameters of the H-shaped microrobot model.}
\label{tab:geometry}
\begin{tabular*}{\textwidth}{@{\extracolsep\fill}llll}
\toprule
Parameter & Symbol & Value & Units \\
\midrule
\multicolumn{4}{l}{\textit{Arm geometry (full arm)}} \\
Arm length        & $L_{\mathrm{arm}}$  & 5.0 & mm \\
Arm width         & $w$                 & 1.0 & mm \\
Arm thickness     & $t$                 & 1.0 & mm \\
Segments per arm  & $N_{\mathrm{seg}}$  & 5   & --- \\
\midrule
\multicolumn{4}{l}{\textit{Gold rod crossbar}} \\
Rod length & $L_{\mathrm{rod}}$ & 3.0 & mm \\
Rod radius & $r_{\mathrm{rod}}$ & 25  & \textmu m \\
\midrule
\multicolumn{4}{l}{\textit{Contact leg geometry}} \\
Hemisphere radius            & $R_{\mathrm{hemisphere}}$ & 0.15 & mm \\
Anterior leg position    & $x_{\mathrm{ant}}$    & 4.0  & mm \\
Posterior leg position   & $x_{\mathrm{post}}$   & 1.0  & mm \\
Contact legs per arm     & ---                   & 2    & --- \\
Support hemispheres per arm  & ---                   & 3    & --- \\
\midrule

\bottomrule
\end{tabular*}
\end{table}

 \subsection{Friction and Contact Model}
\label{subsec:friction}
 Leg-surface contact is modelled using Hertzian contact mechanics via the Simscape Spatial Contact Force Block. At the nominal static load ($F_n \approx 26.9$~\textmu N per leg), the peak and mean Hertzian contact pressure are $p_0 \approx 10.7$~kPa and $\bar{p} \approx 7.1$~kPa respectively \cite{johnson1985}.

Friction is determined via the regularised Armstrong–Hélouvry model \cite{armstrong1994}, which combines Coulomb, Stribeck, and viscous contributions into a single law:
\begin{equation}
  F_{\mathrm{fric}} = -\tanh\!\left(\frac{v}{v_{\mathrm{reg}}}\right)
  \left[
    \underbrace{\mu_k F_n}_{F_C}
    + \underbrace{(\mu_s - \mu_k)\,F_n\,
        e^{-(v/v_s)^{2}}}_{F_S}
    + \underbrace{\sigma_{\mathrm{visc}}\,|v|}_{F_\nu}
  \right],
  \label{eq:friction}
\end{equation}
where $v_{\mathrm{reg}}$ is a regularisation velocity that removes the discontinuity at $v = 0$, preventing solver instability at micro-scale contact events \cite{schuderer2025}.
 
To replicate the directional friction asymmetry of biological crawlers (earthworm setae, snake ventral scales), the coefficients in Eq. \eqref{eq:friction} are interpolated between forward and backward values using the smooth switching term $s = \tanh(v/v_{\mathrm{reg}})$:
\begin{align}
  \mu_s &= \tfrac{1}{2}(1+s)\,\mu_{s,\mathrm{fwd}}
           + \tfrac{1}{2}(1-s)\,\mu_{s,\mathrm{bwd}}, \label{eq:mu_s_interp}\\
  \mu_k &= \tfrac{1}{2}(1+s)\,\mu_{k,\mathrm{fwd}}
           + \tfrac{1}{2}(1-s)\,\mu_{k,\mathrm{bwd}}. \label{eq:mu_k_interp}
\end{align}

The forward friction coefficients ($\mu_{s,\mathrm{fwd}} = 0.15$, $\mu_{k,\mathrm{fwd}} = 0.10$) represent design targets for a
low-resistance anisotropic surface, consistent with the range of friction coefficients reported for anisotropic biological surfaces \cite{hu2009} and soft elastomers \cite{berthold2023}. Mechanically, these low values are required to achieve peristaltic locomotion and can be achieved by using angled, claw like features to produce directionally dependent contact mechanics through geometric asymmetry generating low resistance when sliding forward and interlocking with the surface to resist backward-slip. These features replicate the function of setae or scales and for comparison, \citet{hu2009} measured a forward coefficient of $\mu_{\mathrm{fwd}} \approx 0.11$ for snake ventral scales on a rough surface, of comparable order to the forward target adopted here.   

The backward coefficients ($\mu_{s,\mathrm{bwd}} = 1.20$, $\mu_{k,\mathrm{bwd}} = 1.00$) are supported by the Hertzian pressure analysis, the computed max pressure 
$p_0 \approx 10.7$~kPa falls below the lower bound of the pressure range characterised by \citet{berthold2023} (\SI{11.79}{\kilo\pascal}). As the kinetic coefficient of friction for soft elastomers increases as contact pressure decreases, operating below this point places the backward coefficients at or above the upper end of the measured range of $\mu_k = 0.87$–$1.09$ \cite{berthold2023}. The coefficients used within the model therefore represent the high-friction limit accessible at the low contact pressure exhibited by the modelled system and prevent backward slip through the gait cycle.

The resulting static anisotropy ratio of 8:1 ($\mu_{s,\mathrm{bwd}}/\mu_{s,\mathrm{fwd}}$) should be interpreted as an upper-bound performance target, it exceeds the anisotropies realised in comparable bio-inspired surfaces. The model is nonetheless formulated to be sufficiently general that these coefficients can be adjusted to represent crawling on different biological surfaces according to the intended application. A lateral friction coefficient of $\mu_{\mathrm{lat}} = 0.8$ governs resistance to sideways sliding and couples the torque asymmetry generated by the steering controller to changes in heading, representing an approximately isotropic Ecoflex baseline for kinetic friction \cite{berthold2023}. All contact and friction parameters are listed in Table~\ref{tab:friction_params}.

\begin{table}[t]
\caption{Contact and friction model parameters.}
\label{tab:friction_params}
\begin{tabular*}{\textwidth}{@{\extracolsep\fill}llll}
\toprule
Parameter & Symbol & Value & Units \\
\midrule
\multicolumn{4}{l}{\textit{Friction coefficients}} \\
Forward static    & $\mu_{s,\mathrm{fwd}}$  & 0.15 & --- \\
Forward kinetic   & $\mu_{k,\mathrm{fwd}}$  & 0.10 & --- \\
Backward static   & $\mu_{s,\mathrm{bwd}}$  & 1.20 & --- \\
Backward kinetic  & $\mu_{k,\mathrm{bwd}}$  & 1.00 & --- \\
Lateral           & $\mu_{\mathrm{lat}}$    & 0.80 & --- \\
\midrule
\multicolumn{4}{l}{\textit{Friction model parameters}} \\
Stribeck velocity       & $v_s$                  & $1.0\times10^{-4}$ & m\,s$^{-1}$ \\
Regularisation velocity & $v_{\mathrm{reg}}$     & $2.0\times10^{-4}$ & m\,s$^{-1}$ \\
Viscous coefficient     & $\sigma_{\mathrm{visc}}$ & 0.01              & N\,s\,m$^{-1}$ \\
\midrule
\multicolumn{4}{l}{\textit{Derived quantities}} \\
Anisotropy ratio  & $\mu_{s,\mathrm{bwd}}/\mu_{s,\mathrm{fwd}}$ & 8.0 & --- \\
Static/kinetic ratio (fwd) & $\mu_{s,\mathrm{fwd}}/\mu_{k,\mathrm{fwd}}$ & 1.5 & --- \\
Static/kinetic ratio (bwd) & $\mu_{s,\mathrm{bwd}}/\mu_{k,\mathrm{bwd}}$ & 1.2 & --- \\

\end{tabular*}
\end{table}
 
\subsection{Actuation Model}
\label{subsec:actuation}
Cardiomyocyte contraction arises from Ca$^{2+}$ triggered crossbridge cycling. Ca$^{2+}$ binds to troponin C shifting tropomyosin away from myosin binding sites on the actin filament permitting crossbridge formation and sarcomere shortening, the biological basis for active tension. The active stress is converted to joint torque (Eq \ref{eq:tau_joint}) bending the Ecoflex arm segments. Phase-offset torque across each arm's joints produces a travelling wave that combined with anisotropic friction ratchet drives net forward locomotion.

The actuation model utilises a prescribed Ca$^{2+}$ transient acting as the input, a standard simplification for NRCM-actuated biohybrid models \cite{land2014, gattoni2016}. The aforementioned prescribed Ca$^{2+}$ is governed by a cosine rise to a systolic peak followed by an exponential decay to a diastolic baseline. This normalised activation waveform $\alpha_{act}$ is mathematically described as the following:
\begin{equation}
  \alpha_{\mathrm{act}} =
  \begin{cases}
    \dfrac{1}{2}\!\left(1 - \cos\!\left(\dfrac{\pi\,t_{\mathrm{local}}}{\tau_{\mathrm{ttp}}}\right)\right)
      & t_{\mathrm{local}} \leq \tau_{\mathrm{ttp}}, \\[8pt]
    \exp\!\left(-\dfrac{t_{\mathrm{local}} - \tau_{\mathrm{ttp}}}{\tau_{\mathrm{decay}}}\right)
      & t_{\mathrm{local}} > \tau_{\mathrm{ttp}},
  \end{cases}
  \label{eq:ca_transient}
\end{equation}

where $\tau_{\mathrm{ttp}}$ is the time-to-peak and $\tau_{\mathrm{decay}}$ is the exponential decay constant, both
phenotype-dependent.  The local time at each joint $i$
\begin{equation}
  t_{\mathrm{local}} = \mathrm{mod}\!\left(t_{\mathrm{sim}} - \phi_i,\,T_{\mathrm{period}}\right), \qquad \phi_i = \frac{x_i}{v_C},
  \label{eq:phase_offset}
\end{equation}
prescribes the posterior-to-anterior travelling wave with joint positions $x_i = [0, 1, 2, 3]$~mm and phase offsets $\phi_i = [0, 0.25, 0.50, 0.75]$~s. Peristaltic locomotion is prescribed via these fixed inter-joint phase offsets of 0.250~s, an approach consistent with geometry-defined sequential activation demonstrated in cardiomyocyte-driven soft robots \cite{park2016}. The last piece of the prescribed transient is the Ca$^{2+}$ input for the crossbridge model:
\begin{equation}
  \mathrm{Ca}_{\mathrm{input}} = \mathrm{Ca}_{\mathrm{diastolic}}
  + \left(\mathrm{Ca}_{\mathrm{systolic}} - 
  \mathrm{Ca}_{\mathrm{diastolic}}\right)\alpha_{\mathrm{act}},
  \label{eq:ca_input}
\end{equation}

with $\mathrm{Ca}_{\mathrm{systolic}} = 1.0$~\textmu M and
$\mathrm{Ca}_{\mathrm{diastolic}} = 0.1$~\textmu M, standard
NRCM values \cite{gattoni2016, koivumaki2011}.

\subsubsection*{Four-state crossbridge model}
Active tension (and therefore torque) is calculated using the Land~et~al. \ (2017) four-state crossbridge model \cite{land2017}, which resolves the crossbridge formation detailed above into four sequential biochemical states: blocked (B), in which tropomyosin occludes the myosin binding site; unbound (U), in which the site is exposed but no myosin head is attached; weakly bound (W), the pre-powerstroke attached state; and strongly bound (S), the force-generating post-powerstroke state. The complete model comprises six ordinary differential equations (ODEs); the two most directly applicable to force generation govern the strongly-bound fraction $S$ and the distortion variables $z_w$ and $z_s$.

\begin{equation}
    \frac{d(\text{Ca\_TRPN})}{dt} = k_{\text{TRPN}} \left( \left(\frac{\text{Ca\_input}}{\text{Ca\_50}}\right)^{n_{\text{TRPN}}}\cdot  (1 - \text{Ca\_TRPN}) - \text{Ca\_TRPN} \right)
\end{equation}

\begin{equation}
    \frac{dB}{dt} = k_b \cdot \min\left(100, \text{safe\_TRPN}^{-\frac{n_{\text{Tm}}}{2}}\right) \cdot U - k_u \cdot \left(\text{safe\_TRPN}^{\frac{n_{\text{Tm}}}{2}}\right) \cdot B
\end{equation}

\begin{equation}
    \frac{dW}{dt} = k_{uw} \cdot U - k_{wu\_ref} \cdot W - k_{ws} \cdot W - g_{wu} \cdot W
\end{equation}

\begin{equation}
    \frac{dS}{dt} = k_{ws} \cdot W - k_{su\_ref} \cdot S - g_{su} \cdot S
\end{equation}

\begin{equation}
    \frac{dz_w}{dt} = A_{\text{val}} \cdot \frac{d\lambda}{dt} - c_w \cdot z_w
\end{equation}

\begin{equation}
    \frac{dz_s}{dt} = A_{\text{val}} \cdot \frac{d\lambda}{dt} - c_s \cdot z_s
\end{equation}

$A_{val}$, $k_{wu,ref}$ and $k_{su,ref}$ are fixed quantities derived once from the parameters given in Table \ref{tab:land_params}:  $A_{\mathrm{val}} = A_{\mathrm{eff}}\,r_s / ((1-r_s)r_w + r_s)$,  $k_{wu,\mathrm{ref}} = k_{uw}(1/r_w - 1) - k_{ws}$, and 
$k_{su,\mathrm{ref}} = k_{ws}\,r_w(1/r_s - 1)$. $k_b$ is similarly derived from $k_u$, $n_{\mathrm{Tm}}$, $r_s$, $r_w$ and the model's half-activation threshold ($\mathrm{perm50} = 0.35$) \cite{land2017}. $c_w$, $c_s$, $g_{wu}$, and $g_{su}$ are evaluated at each timestep from $\phi$, $\gamma_w$, $\gamma_s$ and the crossbridge population and distortion state. The rest of the variables used within the model can be seen in the table below (Table \ref{tab:land_params}).

\begin{table}[t]
    \centering
    \caption{State variables of the Land crossbridge actuation model. $U$ is not an independent state but an algebraic conservation variable, recovered from the four crossbridge fractions and used as an input to the $B$, $W$ and $S$ rate equations.}
    \label{tab:land_states}
    \begin{tabular}{lll}
        \toprule
        \textbf{State variable} & \textbf{Symbol} & \textbf{Description} \\
        \midrule
        Troponin Ca\textsuperscript{2+} binding & $\mathrm{Ca}_{TRPN}$ & Fraction of troponin (TnC) sites with bound Ca\textsuperscript{2+} \\
        Blocked fraction & $B$ & Fraction of regulatory units blocked by tropomyosin \\
        Weakly bound fraction & $W$ & Fraction of attached, pre-powerstroke crossbridges \\
        Strongly bound fraction & $S$ & Fraction of force-generating, post-powerstroke crossbridges \\
        Weak-state distortion & $z_w$ & Mean elastic distortion of $W$-state crossbridges \\
        Strong-state distortion & $z_s$ & Mean elastic distortion of $S$-state crossbridges \\
        \midrule
        Unbound fraction & $U$ & Unblocked, unbound crossbridges: $U = 1 - B - W - S$ \\
        \bottomrule
    \end{tabular}
\end{table}
 
From the set of Land ODEs, macroscopic active stress is computed as:
\begin{equation}
  T_{\mathrm{active}} = \frac{T_{\mathrm{ref}}}{r_s}
    \bigl(S\,(z_s + 1) + W\,z_w\bigr)\cdot h_{\lambda},
  \label{eq:T_active}
\end{equation}

where $r_s = 0.25$ is the steady-state duty ratio and $h_{\lambda}$ is a length-dependent scaling factor encoding the Frank–Starling mechanism. $T_{\mathrm{ref}}$ is set to 9.2~kPa for the ventricular phenotype, matching mean peak systolic stress measured in neonatal rat ventricular myocyte (NRVM) thin films \cite{alford2010}, used here as a calibrated effective parameter rather than the Land (2017) maximum Ca$^{2+}$-saturated isometric tension convention. For the atrial phenotype, applying the measured atrial-to-ventricular tension ratio of 0.51~\cite{nollet2020} gives $T_{\mathrm{ref}} = 4.7$~kPa.
 
To produce a bending moment within the Ecoflex arm, a torque is needed. This torque drives the posterior-to-anterior arching motion previously described and acts as the basis of the peristaltic gait with lateral steering arising from the torque differential experience between the two arms. Active stress is used as an input into the torque equation which yields: 

\begin{equation}
  \tau_{\mathrm{joint}} = \tau_{\mathrm{sign}}\cdot
    \bigl(T_{\mathrm{active}}\cdot A\cdot d\cdot
    \alpha_{\mathrm{act}}^{\gamma_{\mathrm{exp}}}\bigr)
    \cdot g_{\mathrm{arm}},
  \label{eq:tau_joint}
\end{equation}
where $A = 1 \times 10^{-7}$~m$^2$ is the simulated thin-film cross-sectional area~\cite{sun2020}, $d = 0.5$~mm is the joint moment arm (half the segment width), $\tau_{\mathrm{sign}} = [+1, +1, -1, -1]$ encodes the arching deformation shape, the $\alpha_{\mathrm{act}}^{\gamma_{\mathrm{exp}}}$ term suppresses residual tension between actuation cycles, and $g_{\mathrm{arm}}$ is the per-arm gain used by the steering controller (Section~\ref{subsec:steering}).
 
\begin{table}[t]
\caption{Ca$^{2+}$ transient and wave parameters for atrial and
         ventricular phenotypes.}
\label{tab:actuation_params}
\begin{tabular*}{\textwidth}{@{\extracolsep\fill}lllll}
\toprule
Parameter & Symbol & Atrial & Ventricular & Units \\
\midrule
Max reference tension  & $T_{\mathrm{ref}}$           & 4.7  & 9.2  & kPa \\
Time to peak           & $\tau_{\mathrm{ttp}}$         & 52   & 80   & ms \\
Decay constant         & $\tau_{\mathrm{decay}}$       & 135  & 200  & ms \\
Cycle period           & $T_{\mathrm{period}}$         & 1.5  & 1.5  & s \\
Conduction velocity    & $v_C$                         & 4    & 4    & mm\,s$^{-1}$ \\
Diastolic Ca$^{2+}$    & $\mathrm{Ca}_{\mathrm{diastolic}}$ & 0.1 & 0.1 & \textmu M \\
Systolic Ca$^{2+}$     & $\mathrm{Ca}_{\mathrm{systolic}}$  & 1.0 & 1.0 & \textmu M \\
Torque sign vector     & $\tau_{\mathrm{sign}}$ & \multicolumn{2}{c}{$[+1,+1,-1,-1]$} & --- \\
Thin-film area         & $A$       & \multicolumn{2}{c}{$1\times10^{-7}$}  & m$^2$ \\
Moment arm             & $d$       & \multicolumn{2}{c}{0.5}               & mm \\
\bottomrule
\end{tabular*}
\end{table}
 
\begin{table}[t]
\caption{Land (2017) crossbridge model parameters~\cite{land2017}.}
\label{tab:land_params}
\begin{tabular*}{\textwidth}{@{\extracolsep\fill}llll}
\toprule
Parameter & Symbol & Value & Units \\
\midrule
Troponin binding rate        & $k_{\mathrm{TRPN}}$     & 100   & s$^{-1}$ \\
Troponin Hill coefficient    & $n_{\mathrm{TRPN}}$     & 2.0   & --- \\
Tropomyosin Hill coefficient & $n_{\mathrm{Tm}}$       & 5.0   & --- \\
Ca$^{2+}$ sensitivity (ref)  & $\mathrm{Ca}_{T50,\mathrm{ref}}$ & 0.805 & \textmu M \\
U$\to$W transition rate      & $k_{uw}$                & 182   & s$^{-1}$ \\
W$\to$S transition rate      & $k_{ws}$                & 12    & s$^{-1}$ \\
Crossbridge unbinding rate   & $k_u$                   & 1000  & s$^{-1}$ \\
Distortion rate (W state)    & $\gamma_w$              & 615   & s$^{-1}$ \\
Distortion rate (S state)    & $\gamma_s$              & 8.5   & s$^{-1}$ \\
Steady-state duty ratio      & $r_s$                   & 0.25  & --- \\
Crossbridge state ratio      & $r_w$                   & 0.5   & --- \\
Distortion gain              & $A_{\mathrm{eff}}$      & 25    & --- \\
Distortion decay scaling     & $\phi_{\mathrm{decay}}$ & 2.23  & --- \\
Length-dep.\ force           & $\beta_0$               & 2.3   & --- \\
Length-dep.\ Ca sensitivity  & $\beta_1$               & $-2.4$ & --- \\
\bottomrule
\end{tabular*}
\end{table}
 

\subsection{Gradient Steering Controller}
\label{subsec:steering}
 Under hypoxic conditions, reduced oxygen availability impairs Ca$^{2+}$ cycling decreasing the systolic peak and thus reducing the peak contractile force generated by each arm \cite{allen1987}. The model represents this via a gain parameter $g_{arm}$ that scales the joint torque  (Eq. \eqref{eq:tau_joint}) of each arm independently, simulating the effects of chemotaxis. A proportional controller evaluates the linear oxygen field $O_2=O_{2,ref}+\Delta_x x_{leg}$ at the anterior contact leg positions (front left and right) giving the bilateral oxygen signal $\delta = O_{2,R} - O_{2,L}$ with a deadband threshold of $0.005$ to suppress numerical noise. The resultant per-arm gain of this is:

\begin{align}
  g_L &= \min\!\bigl(\max\!\bigl(1 - k_p\,\delta,\;
          g_{\min}\bigr),\; g_{\max}\bigr), \label{eq:gL} \\
  g_R &= \min\!\bigl(\max\!\bigl(1 + k_p\,\delta,\;
          g_{\min}\bigr),\; g_{\max}\bigr), \label{eq:gR}
\end{align}

where $k_p = 3.0$ is the proportional gain. The bounds $[g_{\min}, g_{\max}] = [0.3, 1.0]$ constrain the gain response to a bounded range: $g_{\max}=1.0$ reflects that differing oxygen levels do not enhance contractility beyond the baseline whilst $g_{\min}=0.3$ prevents complete silencing of the arm and resulting collapse of locomotion. Within this range, gain varies proportionally with $\delta$, and as such both arms remain oxygen responsive rather than fixed. When $O_{2,R} > O_{2,L}$ ($\delta > 0$), the right arm gain increases as the left arm gain decreases, generating an asymmetric torque that, in combination with the lateral friction coefficient produces a yaw moment towards the area of lower oxygen concentration.

\begin{table}[t]
\caption{Gradient steering controller parameters.}
\label{tab:steering_params}
\begin{tabular*}{\textwidth}{@{\extracolsep\fill}llll}
\toprule
Parameter & Symbol & Value & Units/description \\
\midrule
Proportional gain      & $k_p$              & 3.0   & Steering sensitivity \\
Deadband threshold     & ---                & 0.005 & $O_2$ error below which steering is inactive \\
Minimum arm gain       & $g_{\min}$         & 0.3   & Lower bound on per-arm Ca$^{2+}$ gain \\
Maximum arm gain       & $g_{\max}$         & 1.0   & Upper bound on per-arm Ca$^{2+}$ gain \\
Steering enable time   & $t_{\mathrm{enable}}$ & 1.0 & s; delay before steering activates \\
$O_2$ reference level  & $O_{2,\mathrm{ref}}$  & 1.0 & Normalised baseline $O_2$ \\
$x$-gradient (right turn) & $\Delta_x$     & $-50$ & Gradient magnitude in $x$-direction \\
$y$-gradient           & $\Delta_y$         & 0     & No gradient in $y$-direction \\
\bottomrule
\end{tabular*}
\end{table}
 
The complete model is parametrised for two cardiomyocyte phenotypes atrial and ventricular which differ in contractile stress and Ca$^{2+}$ transient kinetics; simulation conditions and extracted metrics are defined in the following section.
 
\subsection{Simulation Conditions}\label{sec:sim_conditions}
The simulation protocol was structured around two main variables, cardiomyocyte phenotype (atrial or ventricular) and locomotion mode (straight-line or gradient steering). Atrial and ventricular cells were chosen as they represent the two principal cardiomyocyte phenotypes that are routinely isolated in cardiac culture and offer differing contractile stress and Ca$^{2+}$ transient kinetics; other cardiac cells (e.g. sinoatrial nodal cells) are specialised for pacemaking or conduction rather than the required contractile stress. This simulation design isolates the contribution of each variable independently, yielding four simulation conditions, Atrial Straight~(AS), Ventricular Straight~(VS), Atrial Gradient~(AG) and Ventricular Gradient~(VG). The straight-line condition provides a uniform oxygen environment with no applied gradient removing any steering effect on locomotion and allowing direct comparison between phenotypes. This clear comparison independent of gradient effects establishes a baseline comparison between the phenotypes for comparison to the later gradient results. The gradient conditions introduce a proportional steering controller and quantify how each phenotype responds to an applied hypoxic oxygen gradient. Straight-line conditions set the gradient magnitude to $\Delta_x=0$, disabling the controller, whilst gradient conditions set $\Delta_x=-50$ (Table \ref{tab:sim_params}), a value chosen to produce an unambiguous heading response within the \SI{60}{\second} simulation window. A simulation duration of $t_\mathrm{sim} = 60$\,s was chosen to allow a practical number of actuation cycles per run, with steady-state metrics computed over the window $t>$\SI{5}{\second} to exclude the initial transient settling period.

 From each run the following metrics were extracted: mean locomotion speed (\textmu m\,s\textsuperscript{$-1$}), total displacement after \SI{60}{\second}(mm), mean stride length (\textmu m\,cycle\textsuperscript{$-1$}), stride coefficient of variation (CoV,~\%), mean leg normal force (\textmu N), mechanical power (\textmu W), dimensionless cost of transport ($\mathrm{CoT} = P/(mgv)$), and inter-joint phase lags (s), the measured time delay between adjacent joints reaching peak activation (verifying preservation of the prescribed travelling wave). Gradient runs additionally report final heading angle (\textdegree), mean heading rate (\textdegree\,s\textsuperscript{$-1$}, evaluated over $t > 5$\,s), and accumulated lateral displacement (mm).


\begin{table}[t]
\caption{Simulation parameters for all four conditions. Parameters shared across
phenotypes appear once; phenotype-specific values are listed in separate
columns.}\label{tab:sim_params}
\begin{tabular*}{\textwidth}{@{\extracolsep\fill}lllll@{}}
\toprule
Parameter & Symbol & Atrial & Ventricular & Units \\
\midrule
\multicolumn{5}{l}{\textit{Actuation}} \\[2pt]
Reference stress
    & $T_\mathrm{ref}$
    & 4.7   & 9.2   & kPa \\
Peak Ca\textsuperscript{2+}
    & $[\mathrm{Ca}]_\mathrm{syst}$
    & 1.0   & 1.0   & \textmu M \\
Time to peak
    & $\tau_\mathrm{ttp}$
    & 40    & 80    & ms \\
Relaxation time
    & $\tau_{r}$
    & 100   & 200   & ms \\
Actuation period
    & $T$
    & 1.5   & 1.5   & s \\[6pt]
\multicolumn{5}{l}{\textit{Friction (substrate)}} \\[2pt]
Forward COF (kinetic)
    & $\mu_{k,\mathrm{fwd}}$
    & 0.10  & 0.10  & --- \\
Backward COF (kinetic)
    & $\mu_{k,\mathrm{bwd}}$
    & 1.00  & 1.00  & --- \\
Forward COF (static)
    & $\mu_{s,\mathrm{fwd}}$
    & 0.15  & 0.15  & --- \\
Backward COF (static)
    & $\mu_{s,\mathrm{bwd}}$
    & 1.20  & 1.20  & --- \\[6pt]
\multicolumn{5}{l}{\textit{Simulation protocol}} \\[2pt]
Simulation duration
    & $t_\mathrm{sim}$
    & 60    & 60    & s \\
Gradient gain (straight)
    & $\Delta_{x}$
    & 0     & 0     & --- \\
Gradient gain (steering)
    & $\Delta_{x}$
    & $-50$ & $-50$ & --- \\
\botrule
\end{tabular*}
\end{table}

\section{Results}
\label{sec:results}
\subsection{Straight-Line Baseline Performance}
\label{sec:straight}

This section comprises two distinct simulation runs differentiated by their cardiomyocyte phenotype, Atrial Straight (AS) and Ventricular Straight (VS). The primary distinction lies in the previously mentioned transient dynamics ($\tau_{ttp}, \tau_{decay}$) and peak output tension ($T_{ref}$), with the aim of establishing the intrinsic locomotion capabilities of each phenotype under identical mechanical conditions prior to the introduction of gradient-based steering.

\begin{figure}[h]
    \centering
    \includegraphics[width=0.9\textwidth]{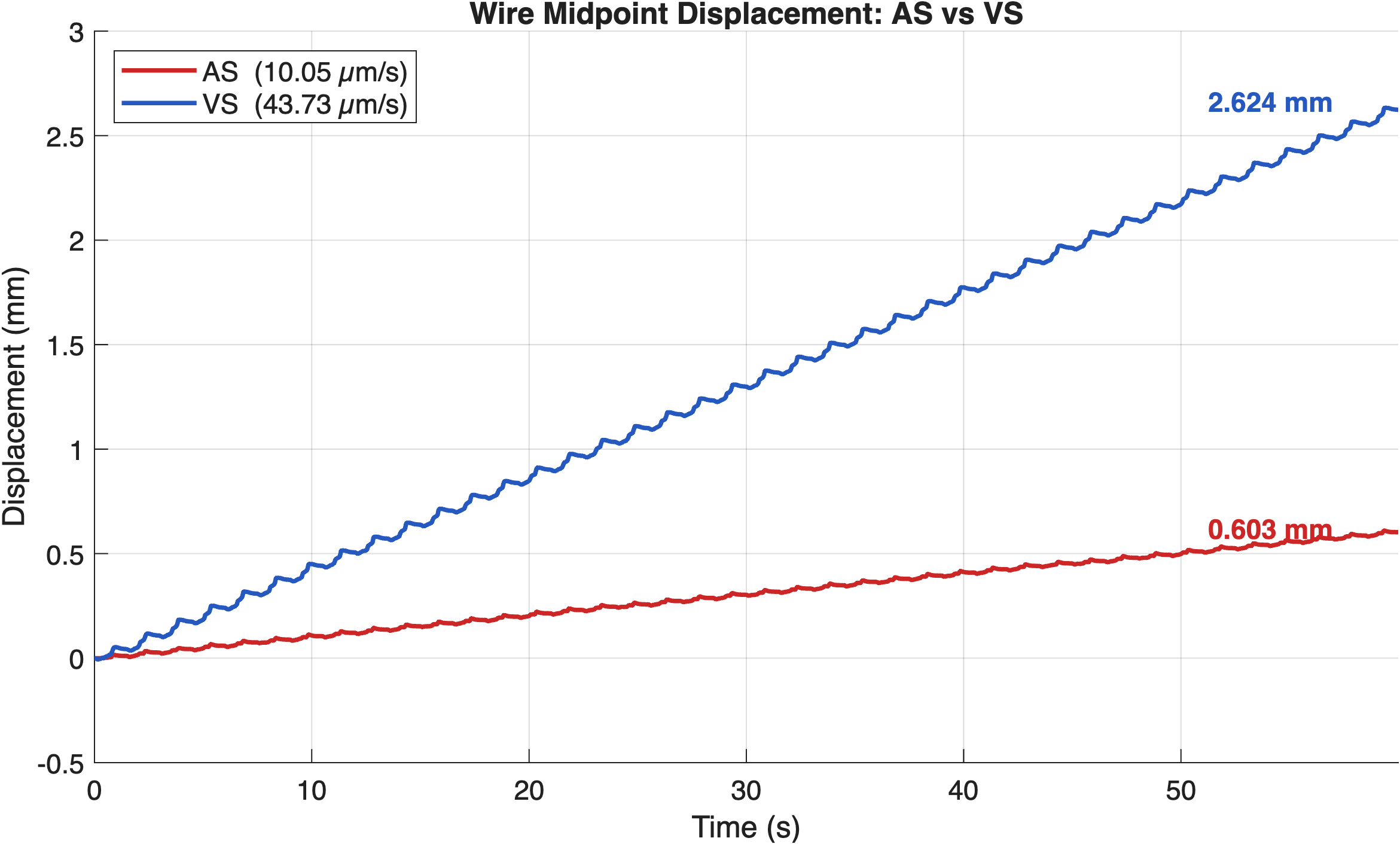}
    \caption{Rod midpoint displacement over the full \SI{60}{\second} simulation
    for AS and VS. The steeper slope of the VS trace reflects the $4.35\times$ greater
    speed  over AS.}
    \label{fig:disp_overlay}
\end{figure}

Both phenotypes produce a stable net-forward displacement over the full 60-second simulation, demonstrating the characteristic staircase shape of peristaltic gait with each step corresponding to one actuation cycle (Figure~\ref{fig:disp_overlay}). All four legs maintain comparable mean normal force (AS: FL 19.3, FR 19.3, BL 19.2, BR 19.4~\textmu N; VS: FL 23.7, FR 24.1, BL 24.0, BR 23.6~\textmu N), varying by no more than \SI{0.2}{\micro\newton} (AS) and \SI{0.5}{\micro\newton} (VS), confirming the geometry is balanced under load for both phenotypes.

The main result from the straight-line runs is the $4.35\times$ speed and displacement ratio of VS over AS that arises from a $1.96\times$ $T_{ref}$ input ratio ($\frac{9.2}{4.7}$). This non-linear force amplification originates from two sources, the Hill force-velocity relationship in which higher contractile stresses produce a higher velocity (seen in the \cite{land2017} model), a direct result of greater ventricular muscular biophysics. The second source is the friction ratchet threshold in which VS significantly exceeds the breakaway friction each cycle whereas AS barely clears it meaning that VS extracts more net displacement per actuation cycle relative to its torque output. The cost of transport ($CoT=\frac{P}{mgv}$ where $P=$ Power, $m=$ mass, $g=$ gravity, $v=$ velocity with lower value indicates more efficiency) comparison of $20.9$ vs $19.5$ shows that the CoT falls within the same range indicating that the energy cost per unit distance is maintained across both phenotypes despite the
$4.05\times$ difference in absolute output power.

\begin{figure}[h]
    \centering
    \includegraphics[width=0.75\textwidth]{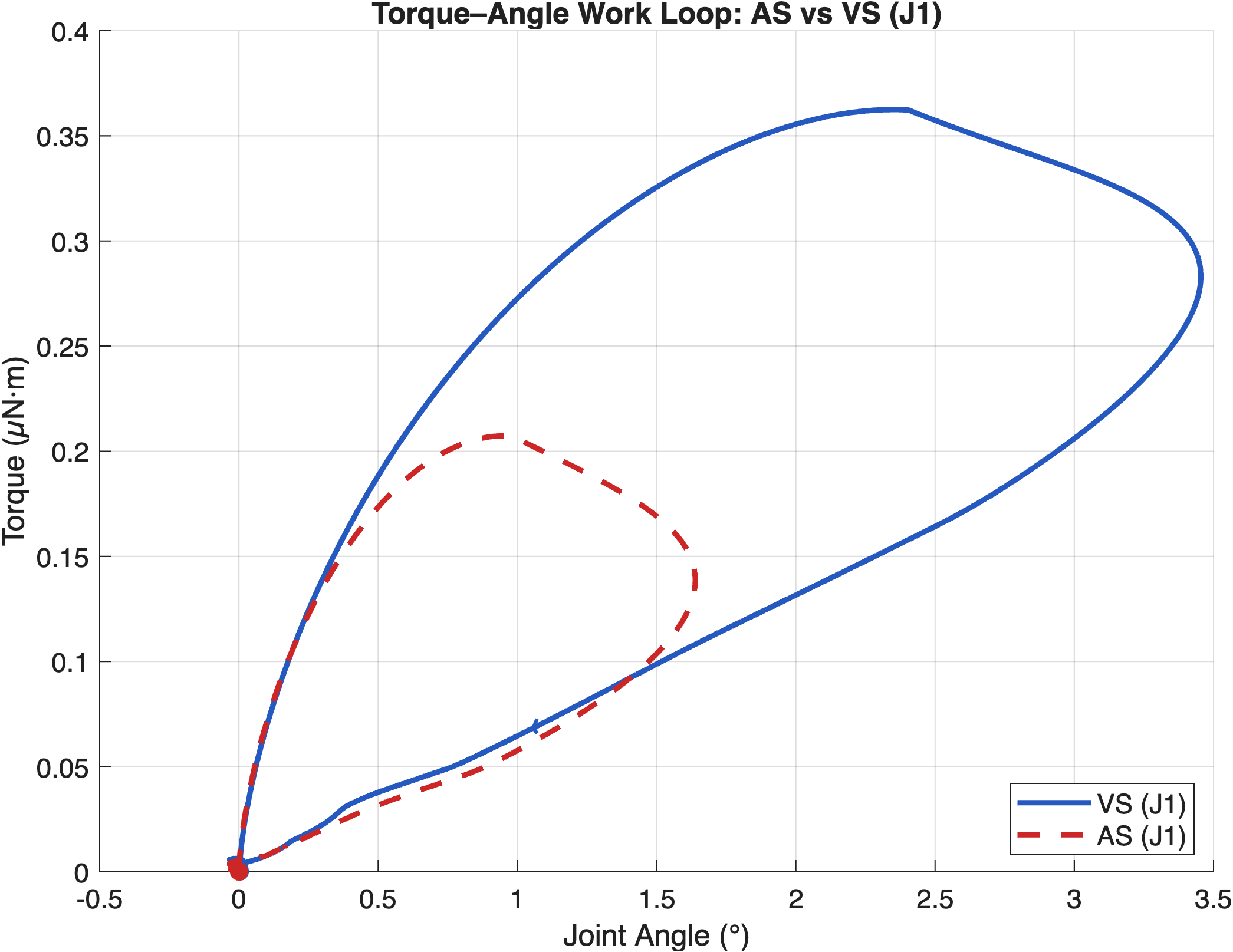}
    \caption{Torque--angle work loop at joint J1 for AS and VS over one
    steady-state cycle ($t = 30$--\SI{31.5}{\second}). The non-zero
    enclosed area is characteristic of a torque--angle hysteresis loop;
    here it reflects net mechanical work performed by the actuator per
    cycle rather than passive material dissipation, with the larger
    area for VS relative to AS reflecting greater mechanical work
    driven by the higher $T_{ref}$.}
    \label{fig:workloop_overlay}
\end{figure}

Torque, stress and angle ratios ($1.75$--$2.43\times$) fall largely within the same range as the $1.96\times$ VS/AS $T_{ref}$ ratio, indicating that the material response scales approximately linearly with the phenotype input. The biggest outlier within the material compliance metrics is the J4 peak angle having a comparison ratio on the higher end at $2.43\times$, this larger deflection angle is consistent with the non-linear relationship between stress-strain exhibited by Ecoflex at higher strain ($1.716\%$ vs $4.179\%$ AS vs VS). The stride coefficient of variation between AS and VS is the biggest variation observed with $8.5\%$ to $1.0\%$ (Table~\ref{tab:4way_comparison}) indicating that the VS stride process is essentially deterministic with an exceptionally consistent gait with little cycle-to-cycle variation in comparison to AS showing meaningful actuation cycle-to-cycle variation. The normal force ratio of $1.24\times$ shows a modest increase in normal force when varying $T_{ref}$ and is expected with VS outputting more force pressing the legs harder into the ground in comparison to AS. The anterior joints (J3, J4) experience a higher strain ($1.58$--$1.71\%$ AS; $4.04$--$4.18\%$ VS) and lower material compliance than the rear joints across both phenotypes (Table~\ref{tab:joint_compact}). The anterior joints engage the mechanical stop during downward actuation and trigger the \cite{land2017} isometric catch mechanism elevating the stress at J3 and J4 relative to the posterior joints. The torque--angle work loop in Figure~\ref{fig:workloop_overlay} illustrates this behaviour, with the larger enclosed area of the VS loop consistent with greater mechanical work per cycle at the joint.

\begin{table}[h]
    \centering
    \caption{Per-joint peak torque, stress, and strain for Atrial Straight (AS) and
    Ventricular Straight (VS). Anterior joints (J3, J4) show consistently higher
    values than posterior joints (J1, J2) for both phenotypes, driven by mechanical
    stop engagement during downward actuation.}
    \label{tab:joint_compact}
    \begin{tabular}{lcccccccc}
        \toprule
        & \multicolumn{4}{c}{\textbf{AS} ($T_{ref}=$\SI{4.7}{\kilo\pascal})}
        & \multicolumn{4}{c}{\textbf{VS} ($T_{ref}=$\SI{9.2}{\kilo\pascal})} \\
        \cmidrule(lr){2-5}\cmidrule(lr){6-9}
        \textbf{Metric} & \textbf{J1} & \textbf{J2} & \textbf{J3} & \textbf{J4}
                        & \textbf{J1} & \textbf{J2} & \textbf{J3} & \textbf{J4} \\
        \midrule
        Peak Torque (\si{\micro\newton\metre})
            & 0.207 & 0.208 & 0.258 & 0.259
            & 0.362 & 0.363 & 0.561 & 0.563 \\
        Peak Stress (\si{\kilo\pascal})
            & 4.147 & 4.164 & 5.156 & 5.182
            & 7.249 & 7.259 & 11.212 & 11.251 \\
        Peak Strain (\%)
            & 1.430 & 1.325 & 1.584 & 1.716
            & 3.015 & 2.904 & 4.037 & 4.179 \\
        \bottomrule
    \end{tabular}
\end{table}

\subsection{Gradient-Based Steering Performance}
\label{sec:gradient}
The gradient-based steering runs apply an oxygen gradient across the simulated environment as described in Section~\ref{subsec:steering} wherein they produce an asymmetric gain generating a yaw moment. A gradient of \texttt{grad\_x}$=-50$ is applied producing a linear oxygen field $O_2(x)=1.0-50x$ (with steering enabled after an initial \SI{1}{\second} settling delay) in which the oxygen concentration drops by $5\%$ per millimetre. This gradient is substantially steeper than the $0.1\%$--$1\%$ per millimetre gradients typically reported across hypoxic tumour margins \cite{hockel2001}; the implications of this gap are addressed below. The key outcomes assessed are the heading change over the simulation time and the retention of forward locomotion metrics relative to the straight-line baseline.

\begin{figure}[h]
    \centering
    \includegraphics[width=\textwidth]{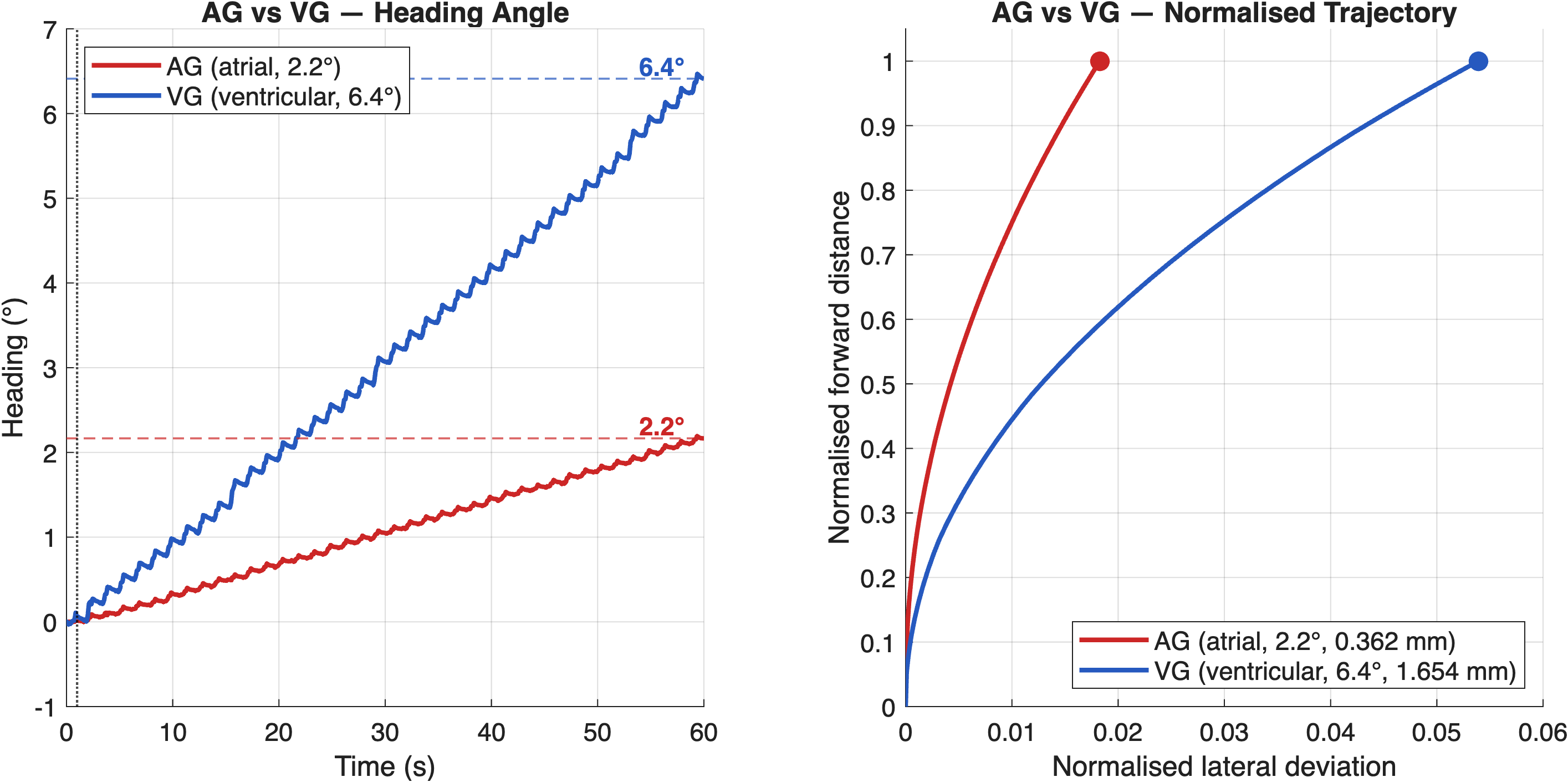}
    \caption{Heading angle (left) and normalised trajectory (right) for AG and VG
over the full \SI{60}{\second} simulation. Trajectories are normalised by each
phenotype's total forward distance (AG: \SI{0.362}{\milli\metre}; VG:
\SI{1.654}{\milli\metre}) so that curvature can be compared independent of
speed. The greater lateral deviation of VG reflects the $2.91\times$ heading
advantage produced by the same gain asymmetry ($g_L/g_R \approx 1.81\times$)
applied to phenotypes with differing absolute torque output.}
    \label{fig:heading_xy_overlay}
\end{figure}

Throughout the simulation (after the \SI{1}{\second} settling delay), the left arm gain saturates immediately to the upper bound ($g_L=1.0$) since the controller's baseline output sits at $g_{\max}$ for any rightward gradient whereas the right arm gain converges to a steady proportional value of $g_R=0.550$ (AG) and $0.553$ (VG) sitting within the operating range opposed to saturating to the lower bound ($g_{\min}$). The resulting gain asymmetry ratios, $\frac{g_L}{g_R}=1.81\times$ (AG) and $1.80\times$ (VG) are near identical between phenotypes and remain constant once reached and as such from that point onwards the system behaves as an open-loop test at a fixed torque asymmetry rather than a continuously responsive closed-loop controller (the resulting heading change can be seen in Figure~\ref{fig:heading_xy_overlay}). As the applied gradient is steeper than a physiological value, the bilateral oxygen signal $\delta$ experienced under a genuine tumour gradient would be proportionally smaller and may fall below the deadband threshold thus requiring further tuning. The results presented should therefore be treated as a proof-of-concept in which the mechanism for steering responds correctly and functions as intended when exposed to a torque asymmetry within the short simulation window. Achieving a reliable heading change rate at physiological gradient magnitudes would require further parameter tuning as outlined in the discussion section.

The two gradient runs experience equal gradient conditions with \texttt{grad\_x}$=-50$ and identical friction and geometry setups allowing for a fair test between phenotypes. The key metric for the gradient runs is the heading which has a ratio of $2.91\times$ (\SI{6.4}{\degree} to \SI{2.2}{\degree}) which arises from near identical gain asymmetry between AG and VG ($1.80\times$ vs $1.81\times$), that is the difference in gain value between the left and right arm. With the same controller input and gradient the different heading is a direct result of VG experiencing a larger torque differential compared with AG. Lateral displacement sees a ratio of $13.1\times$ which is amplified beyond the heading ratio due to VG covering more distance and thus the same angular deviation accrues over a longer path. The key result from the steering standpoint is that both phenotypes confirm that steering physically works and confirms that the model's controller is correctly configured to allow for both forward and lateral displacement whilst still implementing the friction model.

The locomotion achieved with gradient steering enabled is a drop of $\sim$40\% compared with the baseline AS run (Straight-line), with a speed of \SI{6.03}{\micro\metre\per\second} and a distance crawled of \SI{0.362}{\milli\metre}. The final heading achieved for VG is \SI{6.4}{\degree} with a lateral displacement of $\sin(6.4^{\circ})\times1.653\approx$\SI{0.184}{\milli\metre} and a heading rate of \SI{0.107}{\degree\per\second}. This larger offset is explained via the ventricular cells covering more distance per unit time and as such the same angular heading rate accumulates more displacement in the lateral plane.

Locomotion metrics under the gradient exhibit a slightly higher VG/AG speed ratio at $4.57\times$ that of the baseline's $4.35\times$ but remains consistent with phenotype differences operating within the gradient. Torque, stress and angle ratios for the left arm fall within the same range of $1.75$--$2.43\times$ as the baseline where right arm ratios change slightly to the range $1.83$--$2.36\times$. This change in ratio and drop in absolute torque, stress and angle is the expected result due to the scaling of the torque ($\tau_{joint}$) output by $g_{arm}$ and the ratios compared to the baseline hold approximately equal. The stride CoV values at $9.8\%$ (AG) and $11.2\%$ (VG) respectively are comparable in contrast to the $8.5\%$ (AS) to $1.0\%$ (VS) observed in the straight-line runs indicating that the gain asymmetry introduces a sufficient perturbation to reduce the stride consistency exhibited by VS under straight-line conditions. CoT remains comparable at $21.1$ vs $19.4$ indicating that steering does not disproportionately affect either phenotype in terms of efficiency.

Figure~\ref{fig:heading_xy_overlay} shows the heading angle accumulation and normalised trajectory (lateral deviation vs. forward distance) over 60s for both phenotypes.

\subsection{Four-Way Performance Summary}
\label{sec:fourway}

Table~\ref{tab:4way_comparison} presents all four simulation conditions together allowing for the comparison of cross-condition trends to be identified beyond the
pairwise analysis above.

\begin{table*}[h]
    \centering
    \caption{Four-way comparison of key locomotion metrics across all simulation conditions: Atrial Straight (AS), Ventricular Straight (VS), Atrial Gradient (AG), and Ventricular Gradient (VG). Heading is reported only
    for gradient runs.}
    \label{tab:4way_comparison}
    \begin{tabular}{lllll}
        \toprule
        \textbf{Parameter} & \textbf{AS} & \textbf{VS} &
        \textbf{AG} & \textbf{VG} \\
        \midrule
        Speed (\si{\micro\metre\per\second})  & 10.05 & 43.73 & 6.03  & 27.55 \\
        Distance (\si{\milli\metre})          & 0.603 & 2.624 & 0.362 & 1.653 \\
        Stride (\si{\micro\metre\per cycle})  & 15.16 & 66.17 & 8.98  & 41.23 \\
        Stride CoV (\%)                       & 8.5   & 1.0   & 9.8   & 11.2  \\
        Mean $F_n$ (\si{\micro\newton})       & 19.3  & 23.9  & 17.9  & 22.7  \\
        Power (\si{\micro\watt})              & 0.022 & 0.089 & 0.013 & 0.056 \\
        CoT                                   & 20.9  & 19.5  & 21.1  & 19.4  \\
        Phase Lag J1--J2 (\si{\second})       & 0.250 & 0.250 & 0.250 & 0.250 \\
        Phase Lag J2--J3 (\si{\second})       & 0.251 & 0.251 & 0.251 & 0.251 \\
        Phase Lag J3--J4 (\si{\second})       & 0.250 & 0.250 & 0.250 & 0.250 \\
        Final Heading ($^\circ$)              & ---   & ---   & 2.2   & 6.4   \\
        \botrule
    \end{tabular}
\end{table*}

Two metrics remain either totally invariant or approximately invariant across each of the four runs. Phase lags between the joints are totally invariant across each run, indicating that the travelling wave timing is robust to both phenotype and gradient changes. This confirms that conduction velocity and joint spacing govern the wave characteristics not the actuation magnitude. This suggests the peristaltic gait is robust by design and would remain invariant under phenotype changes. CoT spans $19.4$--$21.1$ across each of the four runs indicating that the efficiency per unit distance remains comparable regardless of condition or phenotype and therefore efficiency is not the differentiating factor between phenotype, raw mechanical output ($T_{ref}$) is. Both of these two metrics remaining either wholly or partially invariant across runs indicates that the gait architecture is mechanically stable and the model produces consistent wave propagation and energy conversion regardless of biological input.

Referring to Table~\ref{tab:4way_comparison} one phenotype dominance emerges, ventricular. Across both straight-line and gradient conditions the ventricular speed advantage of $4.35\times$ (VS) and $4.57\times$ (VG) is preserved, confirming that the phenotype is the dominant factor in locomotion performance and that the gradient does not fundamentally alter the phenotype speed relationship. This performance differential confirms that for peak locomotion performance ventricular cardiomyocytes emerge as the logical choice. Both phenotypes confirm gradient-based steering, with ventricular cells producing $2.91\times$ greater heading change than atrial cells under identical conditions. The physical significance of these phenotype differences and their implications for robot design are discussed in the following section.

\section{Discussion}
\label{sec:discussion}

\subsection{Phenotype Comparison}
The principal result of the model is that ventricular cells produce a $4.35\times$ increase in locomotion speed over atrial cells at an equivalent cost of transport, arising from a $1.96\times$ contractile stress input ratio ($T_{ref}$, VS/AS). This leads to ventricular cells being the logical design choice based on model results which is supported by a two-sided argument, ventricular cells are not just better, atrial cells are actively disadvantaged due to their proximity to the friction threshold. The $4.35\times$ speed differential represents an upper bound of Hill amplification as a result of the idealised friction conditions (low forward friction coefficients), under realistic conditions the amplification would persist but would be expected to compress toward the input stress ratio ($1.96\times$) as the friction-ratchet contribution diminishes. 

This result is further reinforced by the CoT due to the $4.35\times$ faster ventricular cells sharing a similar CoT to atrial cells ($19.5$ to $20.9$). This represents another design implication as ventricular cells are significantly faster whilst remaining at approximately the same energy cost per unit distance and as such for a robot operating in a constrained environment, energy expenditure becomes a meaningful consideration. Essentially, ventricular cells represent a speed advantage at little additional energy cost over atrial cells further strengthening the case for ventricular cell selection. The model's results sit within the appropriate range in comparison to similar robots. \citet{sun2020} reports a speed of \SI{10}{\micro\meter\per\second} which is consistent with the atrial results whilst also highlighting the meaningful improvement that the ventricular runs make over the existing devices. As the limitations below highlight, this result is under ideal friction conditions as opposed to the rough substrate used by \citet{sun2020} however, the anisotropy ratio used within the model ($8:1$) is higher than the realistic range of $1.6--2.4\times$ reported by \citet{manoonpond2016} for a similar material; if reduced to this range, the model would be expected to yield speed results similar to the aforementioned paper.

The conclusion regarding phenotype is that, whilst the idealised friction and Ca$^{2+}$ transient reduce the model's stochastic realism (low cycle-to-cycle variability), the phenotype-driven amplification itself arises from the Hill force-velocity relationship acting on a genuine biophysical input difference ($T_{ref}$), a mechanism independent of the model's idealisations. The resulting trend between atrial and ventricular cells therefore represents a meaningful and justified design conclusion, even if the absolute values reported here are upper bounds.

\subsection{Steering}
Both phenotypes steer and produce a measurable heading offset, confirming the gradient-based steering as a viable locomotion control mechanism. This is especially meaningful as the robot could self-direct towards the area it needs to travel to, exhibiting a degree of autonomous behaviour, identified as one of the main challenges for biohybrid robots \cite{webster2023biohybrid}. The proportional steering controller works effectively in both phenotypes and produces the desired offset without collapsing forward locomotion in either.

The $2.91\times$ VG/AG heading difference is a further result to support ventricular being the better design choice. As the controller is the same between phenotypes, the difference in heading is a direct result of the mechanical differences between the two cell types with the same non-linear amplification applying to the yaw moment. The ventricular cells generate a larger yaw per cycle which represents the same phenotype amplification effect as outlined previously in the straight-line runs. As described in Results, the asymmetric steady-state gain response makes this effectively an open-loop test at a fixed torque asymmetry. The present results therefore confirm the steering mechanism functions correctly rather than demonstrating performance under physiological conditions, illustrating that higher biological fidelity would require further controller development and parameter tuning. The gradient-based simulation runs therefore serve as a proof-of-concept of the mechanism, not a demonstration of closed-loop navigation.

\subsection{Limitations}
Several modelling simplifications bound the fidelity of these results. 

The lumped-parameter discretisation of the arms into five solids linked by revolute joints was chosen for solver and computational stability. The discretisation is valid within the current strain range ($\sim$1.5--5\%), higher $T_{ref}$ values (e.g. more mature cardiomyocytes) would exceed the linear elastic regime assumed by the beam-stiffness formulation. Adhesive forces are neglected as they are negligible  relative to the applied normal and actuation forces at the operating scale, although adhesion forces would become dominant if the robot were scaled toward $\sim$\SI{100}{\micro\meter}~\cite{johnson1985}. The idealised, deterministic friction and Ca$^{2+}$ transient further produce unrealistically low cycle-to-cycle variation (CoV as low as $1.0\%$ for VS), a simplification not expected in a physical system.

A further limitation concerns the biological parameterisation. Direct thin-film measurements exist only for mature ventricular myocytes \cite{alford2010}, atrial $T_{ref}$, $\tau_{ttp}$ and $\tau_{decay}$ are derived from the adult rat atrial-to-ventricular ratios given by \citet{nollet2020} and as such introduce a cross-age uncertainty that would most affect the atrial phenotype given its proximity to the friction threshold.

\section{Conclusion}
\label{sec:conclusion}

This paper establishes, to our knowledge, the first computational model to treat cardiomyocyte phenotype as an explicit design variable for biohybrid locomotion and the first to demonstrate autonomous oxygen-gradient steering in a cardiomyocyte-based platform. The final model comprises the robot's geometry defined within Simscape Multibody with an actuation model based upon that proposed initially by \citet{land2017}, the regularised Armstrong--H\'{e}louvry friction model and proportional oxygen-gradient steering. Four simulation conditions were run (AS, VS, AG, VG) across two different phenotypes and two gradient conditions, establishing results covering locomotion mechanics, phenotype comparison and gradient-steering.

The model successfully produces peristaltic crawling via the friction ratchet mechanism. Both phenotypes produce net-forward displacement, confirming that the actuation and anisotropic friction model is viable at the microscale. The central finding is the $4.35\times$ speed amplification from a $1.96\times$ stress input, leading to the conclusion that the ventricular phenotype is the logical design choice based upon the evidence of the model. The ventricular phenotype is faster and equivalently efficient, producing a result that sits within the range of comparable biohybrid devices \cite{alford2010, sun2020}. The proof-of-concept gradient-based steering is established, highlighting that the robot's mechanism can withstand the yaw moment and produce a measurable heading offset, which serves as a basis for a full spatially varying gradient.

Treating cardiomyocyte phenotype as an explicit design parameter rather than a fixed biological value is an underexploited lever in biohybrid robot design. This work shows that phenotype selection alone can yield a $4.35\times$ change in locomotion speed at an equivalent cost of transport, combined with the demonstrated capacity for autonomous, self-guided steering, the findings support the long term feasibility of cardiomyocyte driven microrobots. The use case for the proposed biohybrid robot extends beyond solid tumours and other cancerous sites to any hypoxic region or ischaemic tissue and would reduce systemic drug exposure relative to conventional treatments.  As biohybrid robots mature towards clinical relevance, phenotype selection and self-guided steering offer biologically grounded design variables that are currently underexploited within the field and could considerably increase their applicability and impact. Cardiomyocytes' capacity for autonomous actuation and steering without external stimuli or added control complexity, positions phenotype as a design parameter ready for adoption in future biohybrid platforms.


\section*{Declarations}
\section*{Funding} 
M.E.G. is funded by the Somabotics project, grant number EP/Z534808/1. 
C.V.G. has received funding from the BHF, UK (RG/18/8/33673)
\section*{Conflict of interest} We have no conflict of interest to declare. 
\section*{Code and data availability} 
Code files available at https://github.com/NLilly672/Biobot
Data can be found at: https://doi.org/10.5281/zenodo.21454755.

\section*{Author contribution} 
N.L.: Methodology, Software, Validation, Formal analysis, Investigation, Writing, Original Draft.
G.B., C.V.G., M.E.G.: Conceptualization, Writing, Review \& Editing, Supervision.







\bibliography{sn-bibliography} 

\end{document}